\documentclass[12pt]{article}
\usepackage{a4}
\usepackage{amsfonts}
\usepackage{amsmath,amssymb}

\usepackage{graphicx}

\def\un#1{\relax\ifmmode\@@underline#1\else
        $\@@underline{\hbox{#1}}$\relax\fi}

\let\du=\du                     

\def\a{\alpha}

\def\c{\chi}
\def\d{\delta}

\def\f{\phi}
\def\g{\gamma}

\def\j{\psi}

\def\l{\lambda}
\def\m{\mu}
\def\n{\nu}
\def\o{\omega}

\def\q{\theta}

\def\s{\sigma}

\def\x{\xi}

\def\F{\Phi}

\def\O{\Omega}

\def\vf{\varphi}

\def\ce{{\cal E}}

\def\car{{\cal R}}

\def\cz{{\cal Z}}

\def\bo{{\raise-.3ex\hbox{\large$\Box$}}}               
\def\pa{\partial}                                       
\def\TH{{\raise.2ex\hbox{$\displaystyle \bigodot$}\mskip-4.7mu \llap H \;}}
\def\face{{\raise.2ex\hbox{$\displaystyle \bigodot$}\mskip-2.2mu \llap {$\ddot
        \smile$}}}                                      
\def\dg{\sp\dagger}                                     

\def\sp#1{{}^{#1}}                              
\def\Bar#1{\overline{#1}}                       
\def\abs#1{\left| #1\right|}                    
\def\leftrightarrowfill{$\mathsurround=0pt \mathord\leftarrow \mkern-6mu
        \cleaders\hbox{$\mkern-2mu \mathord- \mkern-2mu$}\hfill
        \mkern-6mu \mathord\rightarrow$}
\def\dvec#1{\vbox{\ialign{##\crcr
        \leftrightarrowfill\crcr\noalign{\kern-1pt\nointerlineskip}
        $\hfil\displaystyle{#1}\hfil$\crcr}}}           
\def\dt#1{{\buildrel {\hbox{\LARGE .}} \over {#1}}}     

\def\frac#1#2{{\textstyle{#1\over\vphantom2\smash{\raise.20ex
        \hbox{$\scriptstyle{#2}$}}}}}                   
\def\sfrac#1#2{{\vphantom1\smash{\lower.5ex\hbox{\small$#1$}}\over
        \vphantom1\smash{\raise.4ex\hbox{\small$#2$}}}} 
\def\bfrac#1#2{{\vphantom1\smash{\lower.5ex\hbox{$#1$}}\over
        \vphantom1\smash{\raise.3ex\hbox{$#2$}}}}       
\def\afrac#1#2{{\vphantom1\smash{\lower.5ex\hbox{$#1$}}\over#2}}    

\def\[{\lfloor{\hskip 0.35pt}\!\!\!\lceil}
\def\]{\rfloor{\hskip 0.35pt}\!\!\!\rceil}
\def\Lag{{\cal L}}
\def\du#1#2{_{#1}{}^{#2}}

\def\fracm#1#2{\hbox{\large{${\frac{{#1}}{{#2}}}$}}}
\def\ha{{\fracmm12}}

\def\un{\underline}
\def\fracmm#1#2{{{#1}\over{#2}}}

\def\low#1{{\raise -3pt\hbox{${\hskip 0.75pt}\!_{#1}$}}}

\def\Dot#1{\buildrel{_{_{\hskip 0.01in}\bullet}}\over{#1}}
\def\dt#1{\Dot{#1}}

\newskip\humongous \humongous=0pt plus 1000pt minus 1000pt
\def\caja{\mathsurround=0pt}
\def\eqalign#1{\,\vcenter{\openup2\jot \caja
        \ialign{\strut \hfil$\displaystyle{##}$&$
        \displaystyle{{}##}$\hfil\crcr#1\crcr}}\,}
\newif\ifdtup

\newcommand{\be}{\begin{equation}}
\newcommand{\ee}{\end{equation}}
\newcommand{\nbe}{\begin{equation*}}
\newcommand{\nee}{\end{equation*}}

\newcommand{\lb}{\label}

\begin{document}

\thispagestyle{empty}

{\hbox to\hsize{
\vbox{\noindent RESCEU-8/11 \hfill July 2012 \\  IPMU11-0196}}}

\noindent
\vskip2.0cm
\begin{center}

{\large\bf Inflation and Nonminimal Scalar-Curvature Coupling 
                     in Gravity and Supergravity}

\vglue.3in

Sergei V. Ketov~${}^{a,b}$ and Alexei A. Starobinsky~${}^{c,d}$ 
\vglue.1in

${}^a$~Department of Physics, Tokyo Metropolitan University, 
Minami-ohsawa 1-1, Hachioji-shi, Tokyo 192-0397, Japan \\
${}^b$~Kavli Institute for the Physics and Mathematics of the Universe (IPMU), The 
           University of Tokyo, Chiba 277-8568, Japan \\
${}^c$~Landau Institute for Theoretical Physics RAS, Moscow 119334, Russia\\
${}^d$~Research Center for the Early Universe (RESCEU), Graduate School of 
Science, The University of Tokyo, Tokyo 113-0033, Japan

\vglue.1in
ketov@tmu.ac.jp, alstar@landau.ac.ru
\end{center}

\vglue.3in

\begin{center}
{\Large\bf Abstract}
\end{center}
\vglue.1in

Inflationary slow-roll dynamics in Einstein gravity with a nonminimal
scalar-curvature coupling can be equivalent to that in the certain $f(R)$ 
gravity theory. We review the correspondence and extend it to N=1 supergravity. 
The nonminimal coupling in supergravity is rewritten in terms of the standard 
(`minimal') N=1 matter-coupled supergravity by using curved superspace. The established
 equivalence between two different inflationary theories means {\it the same} inflaton 
scalar potential, and does not imply the same post-inflationary dynamics and reheating.

\newpage

\section{Introduction}

The standard mechanism of inflation in field theory uses the Einstein gravity 
minimally coupled to a non-gravitational scalar field (called {\it inflaton}) 
whose potential energy drives inflation. The inflaton scalar potential should 
be flat enough to meet the slow-roll conditions during the inflationary period.
This is not the only possibility since inflaton may have a purely geometrical
 origin and then gravity has to be modified also. However, as we argue below, 
both alternatives represent the particular cases of the more general situation
 when inflaton field is non-minimally coupled to gravity that leads to a 
variant of scalar-tensor gravity. Great success of inflationary cosmology over
 the recent past, which has led to the definitive predictions about the 
post-inflationary Universe confirmed by observations, has left unanswered the 
fundamental question about the {\it origin} of inflaton particle and its 
scalar potential. Knowing the inflaton origin would fix inflaton 
{\it interactions} with other particles. In turn, it would lead to definitive 
physical predictions about reheating after inflation and the origin of all 
known elementary particles from inflaton decay. 

One may pursue two different strategies in a theoretical search for inflaton.
Inflaton may be either a new exotic particle or something that we already know
 `just around the corner'. We adopt the second (`economical') approach. In this
paper we consider the two natural possibilities: Higgs inflation and $(R+R^2)$ 
inflation, where $R$ is the Ricci scalar. In the Higgs inflation, inflaton is 
identified with Higgs particle nonminimally coupled to gravity \cite{bs} (see 
also \cite{BKS08,BMS09,SHW09,BKSS09} for the study of the Standard Model loop 
corrections to the Higgs scalar potential and their effect on observational 
quantities). In the $(R+R^2)$ inflation \cite{star, star2} (see also 
\cite{vil85}), which historically was the first fully developed inflationary 
model having a graceful exit from the initial de Sitter stage to the final 
radiation dominated Friedmann-Robertson-Walker (FRW) stage through an 
intermediate period of matter creation and heating, the inflaton field 
(dubbed scalaron) is spin-$0$ part of spacetime metric, so it has the purely 
geometrical origin. 
 
Because inflationary dynamics of the universe is fully determined by 
the effective inflaton potential, some apparently very different models of 
inflation may lead to the same inflationary physics. It occurs in the case 
under our consideration also. Indeed, it was already noticed in \cite{BKS08} 
that the spectra of scalar and tensor perturbations generated in the 
$(R+R^2)$-inflationary model \cite{mukh} are the {\em same} as those in the
Higgs inflationary model with the tree level Higgs potential \cite{bs} in the 
limit of the infinitely large coupling $\xi\to\infty$ (see also \cite{bg1}). 
This equivalence of observational predictions is the consequence of the 
asymptotic duality of both models. Namely, the inflaton scalar potential, 
derived from the nonminimal coupling \cite{bs}, does coincide with the 
effective inflaton scalar potential that follows 

\newpage

\noindent from the $(R+R^2)$-inflationary model \cite{star,star2} in the limit 
$\xi\to\infty$.~\footnote{This duality also provides a novel microscopic 
mechanism of generating macroscopic $f(R)$ gravity which, in contrast to the
one-loop quantum gravitational effects, does {\em not} lead to the appearance 
of the extra $R_{\m\n}R^{\m\n}$ term producing undesirable ghost or tachyon 
massive gravitons at the quasi-classical level. However, {\em not any} $f(R)$ 
gravity theory can be obtained this way.} The main topic of our paper is 
upgrade of this duality to N=1 supergravity. 

The established duality of both models is valid during inflation and even for 
some time after it, but not for the whole post-inflationary evolution. 
Reheating mechanisms in those models are different too. As a result, the 
{\it reheating} temperature $T_{\rm reh}$ after Higgs inflation is about 
$10^{13}~GeV$ \cite{bs}, whereas after $(R+R^2)$ inflation one finds 
$T_{\rm reh}\approx 10^9~GeV$ \cite{star2}, see also \cite{gorb}.

The motivation of \cite{bs} was based on the `most minimal' assumption that 
there is no new physics beyond the Standard Model up to the Planck scale. The 
most economical mechanism of inflation can be based on a new (beyond the 
Standard Model) Higgs dynamics instead of introducing a new particle. We 
assume the new physics beyond the Standard Model, which is given by 
supersymmetry. Then it is quite natural to search for the most economical 
mechanism of inflation in the context of supergravity. And we are not forced 
to identify inflaton with a Higgs particle of the Minimal Supersymmetric 
Standard Model.

Our paper is organized as follows. In sec.~2 we briefly review chaotic 
inflation with nonminimal coupling to gravity \cite{bs} (see also the old paper
\cite{spok} albait without a connection to Higgs scalar). In sec.~3 we briefly 
review chaotic inflation in $(R+R^2)$ gravity \cite{star,star2}, and 
prove its duality to that of sec.~2. In sec.~4 we outline a construction 
of the new supergravity theory proposed in ref.~\cite{us} and called $F(\car)$ 
supergravity. Our main results are given in secs.~5 and 6. Our conclusions are
in sec.~7.

\section{Inflation with nonminimal coupling to gravity}

Consider the 4D Lagrangian 
\be \lb{bl}
\Lag_{\rm J} = \sqrt{-g_{\rm J}}\left\{ -\ha (1+\x\f_{\rm J}^2)R_{\rm J} + 
\ha g^{\m\n}_{\rm J}\pa_{\m}\f_{\rm J}\pa_{\n}\f_{\rm J} -V(\f_{\rm J}) \right\}
\ee
where we have introduced the real scalar field $\f_{\rm J}(x)$, nonminimally
coupled to gravity (with the coupling constant $\x$) in the Jordan frame, with the 
Higgs-like scalar potential  
\be \lb{hpot}
V(\f_{\rm J}) = \fracmm{\l}{4}(\f_{\rm J}^2-v^2)^2 
\ee
We use the units $\hbar=c=M_{\rm Pl}=1$, where $M_{\rm Pl}$ is the reduced Planck 
mass,  with the spacetime signature $(+,-,-,-)$.

The action (\ref{bl}) can be rewritten to Einstein frame by redefining the 
metric via a Weyl transformation,
\be \lb{me}
g^{\m\n} = \fracmm {g_{\rm J}^{\m\n}}{1+\x\f_{\rm J}^2}
\ee
It gives rise to the standard Einstein-Hilbert term $(-\ha R)$ for gravity in
the Lagrangian. However, it also leads to a nonminimal (or noncanonical) 
kinetic term of the scalar field $\f_{\rm J}$. To get the canonical kinetic term,
a scalar field redefinition is needed, $\f_{\rm J}\to \vf(\f_{\rm J})$, 
subject  to the condition
\be \lb{fre}
 \fracmm{d\vf}{d\f_{\rm J}} = \fracmm{ 
\sqrt{1+\x(1+6\x)\f_{\rm J}^2}}{1+\x\f_{\rm J}^2}
\ee
As a result, the non-minimal theory (\ref{bl}) is classically equivalent to
the standard (canonical) theory of the scalar field $\vf(x)$ minimally coupled 
to gravity,
\be \lb{mina}
\Lag_{\rm E} = \sqrt{-g}\left\{ -\ha R + \ha g^{\m\n}\pa_{\m}\vf\pa_{\n}\vf
 -V(\vf) \right\}
\ee
with the scalar potential 
\be \lb{hpote}
V(\vf) = \fracmm{V(\f_{\rm J}(\vf))}{[1+\x\f_{\rm J}^2(\vf)]^2}
\ee

Given a large positive $\x\gg 1$, in the small field limit one finds from
eq.~(\ref{fre}) that $\f_{\rm J}\approx \vf$, whereas in the large $\vf$ limit one
gets
\be \lb{fres}
 \vf \approx \sqrt{\fracmm{3}{2}} \log (1+\x\f_{\rm J}^2) 
\ee

Then eq.~(\ref{hpote}) yields the scalar potential:\\
(i) in the {\it very small} field limit, $v \ll\vf < 
\sqrt{\fracmm{2}{3}}\x^{-1}$, as 
\be \lb{vsp}
V_{\rm vs}(\vf) \approx   \fracmm{\l}{4}\vf^4
\ee
(ii) in the {\it small} field limit, $\sqrt{\fracmm{2}{3}}\x^{-1}<\vf\ll
\sqrt{\fracmm{3}{2}}$, as
\be \lb{smp}
V_{\rm s}(\vf) \approx   \fracmm{\l}{6\x^2}\vf^2,
\ee
(iii) and in the {\it large} field limit, $
\vf\gg\sqrt{\fracmm{2}{3}}\x^{-1}$, as 
\be \lb{lp}
V(\vf) \approx \fracmm{\l}{4\x^2}\left(
1-\exp\left[ -\sqrt{\fracmm{2}{3}} \vf\right] \right)^2
\ee
We have assumed here that $\x\gg 1$ and $v\x\ll 1$. 

It was proposed in ref.~\cite{bs} to identify inflaton with Higgs particle, which
requires the parameter $v$ to be the order of weak scale, and the coupling 
$\l$ be the Higgs boson selfcoupling at the inflationary scale. The scalar 
potential (\ref{lp}) is well known to be perfectly suitable to support a 
slow-roll inflation, while its consistency with the COBE normalization 
condition for the observed CMB amplitude of density perturbations (eg., at the 
e-foldings number $N_e=50\div 60$) gives rise to $\x/\sqrt{\l}\approx 5\cdot 10^4$  
\cite{bs}. The scalar potential (\ref{smp}) corresponds to the post-inflationary 
matter-dominated epoch described by the oscillating inflaton field $\vf$ with the 
frequency
\be \lb{freq}
\o = \sqrt{\fracmm{\l}{3}}\,\x^{-1}
\ee  

\section{Inflation in $(R+R^2)$ gravity}

It has been known for a long time \cite{star,star2} that viable inflationary 
models can be easily constructed in (non-supersymmetric) $f(R)$-gravity 
theories (see eg., refs.~\cite{fgrev} for a recent review) with the action
\be \lb{fgr} S= 
\int d^4x \sqrt{-g}~ f(R)
\ee
whose function $f(R)$ begins with the Einstein-Hilbert term, $(-\ha R)$, while the 
rest takes the form $R^2C(R)$ for $R\to\infty$, with a slowly varying function 
 $C(R)$. The simplest model is given by $C(R)=const.\neq 0$ with
\be \lb{star}
f(R) = -\ha \left(R-\fracmm{R^2}{6M^2}\right) 
\ee
The theory (\ref{star}) is known as the excellent model of chaotic inflation 
\cite{linde,mukh,kan}. The coefficient in front of the second term on the 
right-hand-side of eq.~(\ref{star}) is chosen so that $M$ actually coincides 
with the rest mass of the scalar particle (scalaron) appearing 
in $f(R)$-gravity at low 
curvatures $\abs{R}\ll M^2$ or in flat spacetime, in particular. The model 
fits the observed amplitude of scalar perturbations if  
$M/M_{\rm Pl} \approx 1.5 \cdot  10^{-5} 
(50/N_e)$, and gives rise to the spectral index  
$n_s-1\approx -2/N_e\approx -0.04(50/N_e)$ 
and the scalar-to-tensor ratio $r\approx 12/N_e^2 \approx 0.005(50/N_e)^2$, 
in terms of the e-foldings number $N_{e}\approx (50\div 55)$ depending upon 
details of reheating after inflation \cite{star2,gorb,kan}. Despite of the fact that 
it has been known for more than 30 years, the model (\ref{star}) remains viable 
and is in agreement with the most recent WMAP7  observations of 
$n_s=0.963\pm 0.012$ and $r<0.24$  (with 95\% CL) \cite{wmap}.

As is also well known \cite{oldr}, any $f(R)$ gravity theory
is classically equivalent to the scalar-tensor gravity with the Brans-Dicke 
paremeter $\o_{\rm BD}=0$.  In order to derive the 
corresponding scalar potential, one rewrites the theory (\ref{fgr})
to the equivalent form 
\be \lb{afo}
S_A = \int d^4x \sqrt{-g} \left[ AR - Z(A)\right] 
\ee
where the scalar field  $A$ has been introduced. Via eliminating the 
scalar field $A$ by its algebraic equation of motion from the action 
(\ref{afo}) one gets back the original action (\ref{fgr}) provided that the
functions $f$ and $Z$ are related via Legendre transformation,
\be \lb{leg}
 f(R)= RA(R) -Z(A(R))
\ee 
It follows, in particular, that
\be \lb{lder}
Z'(A)=R \qquad {\rm and}\qquad  f'(R) = A
\ee
where the primes denote the derivatives with respect to the given argument.

A Weyl transformation
\be \lb{we}
 g_{\m\n}\to g_{\m\n}\exp\left[ -\sqrt{\fracmm{2}{3}}\vf\right] 
\ee
with the conformal factor
\be \lb{cf}
\exp\left[ \sqrt{\fracmm{2}{3}}\vf\right] =A
\ee
allows one to bring the action (\ref{afo}) to Einstein frame with the canonical
kinetic terms,
\be \lb{stgr}
S_{\vf} =  \int d^4x\, \sqrt{-g}\left\{ -\fracmm{1}{2} R
+\fracmm{1}{2}g^{\m\n}\pa_{\m}\vf\pa_{\n}\vf + \fracmm{1}{2}
\exp \left[ \fracmm{-4\vf}{\sqrt{6}}\right] Z(A(\vf)) \right\} 
\ee
in terms of the physical (and canonically normalized) scalar field $\vf$, with
the scalar potential
\be \lb{poten}
V(\vf) = -\fracmm{1}{2}\exp \left[
 \fracmm{-4\vf}{\sqrt{6}}\right]
Z\left( \exp \left[ \sqrt{ \fracmm{2}{3} }\vf  \right] \right)
\ee

In the special case \cite{star,star2}
\be \lb{ourc}
f_{\rm S}(R)  =-\ha \left( R -\fracmm{1}{6M^2}R^2 \right) 
\ee
one finds 
\be  \lb{spot}  
V(\vf) =  \fracmm{3}{4}M^2 \left(
1-\exp\left[ -\sqrt{\fracmm{2}{3}} \vf\right] \right)^2
\ee
This inflaton scalar potential is {\it the same} as the one in eq.~(\ref{lp})
provided that we identify the couplings as
\be \lb{ind1}
3M^2 =\fracmm{\l}{\x^2}
\ee
Therefore, we conclude that the physical consequences for inflation in the
model with the nonminimal scalar-curvature coupling (Sec.~2) and the $(R+R^2)$ model
of this section are essentially the same. In particular, the inflaton mass is given
by
\be \lb{infm}
 M = \fracmm{1}{\xi} \sqrt{\fracmm{\l}{3}}
\ee
and is equal to the frequency $(\o)$ of small metric and curvature 
oscillations around the FRW background solutions with low curvature.

\section{Nonminimal coupling in supergravity}

In 4D, N=1 supersymmetry, gravity is to be extended to N=1 supergravity, while a  
scalar field should be complexified and become the leading complex scalar field 
component of a chiral (scalar) matter supermultiplet. In a curved superspace of 
N=1 supergravity, the chiral matter supermultiplet is described by a covariantly
chiral superfield $\F$ obeying the constraint  $\Bar{\nabla}_{\dt{\a}}\F=0$. See
the textbooks \cite{ggrs,wb,buk} for more details. We use here the notation of 
Wess and Bagger \cite{wb}. The standard (generic and minimally coupled) 
matter-supergravity action reads in superspace as follows: 
\be \lb{msg}
S_{\rm MSG}= -3\int d^4x d^4\q E^{-1}\exp \left[ 
-\fracmm{1}{3}K(\F,\Bar{\F})\right] +\left\{ 
\int d^4x d^2\q \ce W(\F) +{\rm H.c.} \right\}
\ee
in terms of the K\"ahler potential $K$ and the superpotential $W$ of the chiral 
supermatter, and the full density $E$ and the chiral density $\ce$ of the superspace 
supergravity. It is convenient to introduce the notation
\be \lb{omega}
 \O = -3 \exp \left[ -\fracmm{1}{3}K \right] \quad {\rm or} \quad
K = -3 \ln \left[ -\fracmm{1}{3}\O \right]
\ee

The non-minimal matter-supergravity coupling in superspace reads
\be \lb{nma}
S_{\rm NM}= \int d^4x d^2\q \ce X(\F)\car +{\rm H.c.}
\ee
in terms of the chiral function $X(\F)$ and the N=1 chiral scalar 
supercurvature superfield $\car$ obeying  $\Bar{\nabla}_{\dt{\a}}\car=0$.
In terms of the field components of the superfields the non-minimal action 
(\ref{nma}) is given by 
\be \lb{bnma}
\int d^4x d^2\q \ce X(\F)\car +{\rm H.c.} = -\fracmm{1}{6}\int d^4x \sqrt{-g}
X(\f_c)R + {\rm H.c.} +\ldots
\ee
where the dots stand for the fermionic terms, and $\f_c=\left. \F\right|=
\f+i\c$ is the leading complex scalar field component of the superfield 
$\F$. Given $X(\F)=-\x\F^2$ with the real coupling constant $\x$, we find 
the bosonic contribution 
\be \lb{sbnm}
S_{\rm NM,bos.}=  \fracmm{1}{6}\x\int d^4x \sqrt{-g}
\left( \f^2-\c^2\right) R 
\ee
It is worth noticing that the supersymmetrizable (bosonic) non-minimal coupling reads
$\left[ \f_c^2 +(\f_c^{\dg})^2\right]R$, not $(\f_c^{\dg}\f_c)R$.

Let's introduce the manifestly supersymmetric nonminimal action (in Jordan frame) as
\be \lb{nmact}
S = S_{\rm MSG} + S_{\rm NM} 
\ee

In curved superspace of N=1 supergravity the (Siegel's) chiral integration rule
\be \lb{sieg}
\int d^4x d^2\q \ce \Lag_{\rm ch}  = \int d^4x d^4\q E^{-1}
\fracmm{\Lag_{\rm ch}}{\car} 
\ee
applies to any chiral superfield Lagrangian $\Lag_{\rm ch}$ with
$\Bar{\nabla}_{\dt{\a}}\Lag_{\rm ch}=0$. It is, therefore, possible to 
rewrite eq.~(\ref{nmact}) to the equivalent form
\be \lb{nmeq}
S_{\rm NM} = \int d^4x d^4\q E^{-1} \left[ X(\F) +\Bar{X}(\Bar{\F})\right]
\ee
We conclude that adding $S_{\rm NM}$ to $S_{\rm MSG}$ is equivalent to
the simple change of the $\O$-potential as ({\it cf}. ref.~\cite{jones})
\be \lb{chan}
\O \to \O_{\rm NM} = \O +X(\F) + \bar{X}(\Bar{\F})
\ee
Because of eq.~(\ref{omega}), it amounts to the change of the K\"ahler 
potential as 
\be \lb{kchan}
K_{\rm NM} = -3 \ln \left[ e^{-K/3} -\fracmm{X(\F)+\Bar{X}(\Bar{\F})}{3}
\right]
\ee
The scalar potential in the matter-coupled supergravity (\ref{msg}) is given by 
 \cite{crem}
\be \lb{crem}
V(\f,\bar{\f})=e^G \left[  
\left( \fracmm{\pa^2 G}{\pa\f\pa\bar{\f}}\right)^{-1}\fracmm{\pa G}{\pa\f}
\fracmm{\pa G}{\pa\bar{\f}} -3\right]
\ee
in terms of the {\it single\/} (K\"ahler-gauge-invariant) function
\be \lb{gfun}
G = K +\ln\abs{W}^2
\ee
Hence, in the nonminimal case (\ref{nmact}) we have
\be \lb{gfunnm}
G_{\rm NM} = K_{\rm NM} + \ln\abs{W}^2
\ee

Contrary to the bosonic case, one gets a nontrivial K\"ahler potential
$K_{\rm NM}$, ie. a {\it Non-Linear Sigma-Model} (NLSM) as the kinetic term of
$\f_c=\f+i\c$ (see eg., ref.~\cite{mybook} for more about the NLSM). Since the 
NLSM target space in general has a nonvanishing curvature, no field redefinition 
generically exist that could bring the kinetic term to the free (canonical) 
form with its K\"ahler potential $K_{\rm free}=\Bar{\F}\F$.    

\section{$F(\car)$ supergravity and chaotic inflation}

The textbook (Poincar\'e) supergravity \cite{ggrs,wb,buk} is the N=1 locally 
supersymmetric extension of Einstein gravity, with the Einstein-Hilbert gravitational
 term in the action. The manifestly N=1 locally supersymmetric extension of all 
$f(R)$ gravity theories (\ref{fgr}) was constructed in curved superspace only 
recently \cite{us} and was called $F(\car)$ supergravity.~\footnote{A component field 
construction, by the use of the 4D, N=1 superconformal tensor calculus, 
was proposed in ref.~\cite{cec}.}  

The $F(\car)$ supergravity is  nicely formulated in a chiral 4D, N=1 
superspace where it is defined by the action
\be  \lb{fact}
 S = \int d^4x d^2\q\, \ce F(\car) + {\rm H.c.}
\ee
in terms of a holomorphic function $F(\car)$ of the covariantly-chiral scalar
curvature superfield $\car$, and the chiral superspace density $\ce$. The chiral
$N=1$ superfield $\car$ has the scalar curvature $R$ as the field coefficient at its
$\q^2$-term. The chiral superspace density $\ce$ (in a WZ gauge) reads
\be \lb{cde}
\ce = e \left( 1- 2i\q\s_a\bar{\j}^a +\q^2 B\right) 
\ee
where $e=\sqrt{-g}$, $\j^a$ is gravitino, and $B=S-iP$ is the complex scalar 
auxiliary field (it does not propagate in the theory (\ref{fact}) despite of the
apparent presence of the higher derivatives). The full component structure of the 
action (\ref{fact}) is very complicated. Nevertheless, it is classically equivalent
 to the standard N=1 Poincar\'e supergravity minimally coupled to the chiral 
scalar superfield, via the supersymmetric Legendre-Weyl-K\"ahler transform 
\cite{us}. The chiral scalar superfield is the superconformal mode of a
supervielbein (in Minkowski or AdS vacuum).   

A relation to the $f(R)$-gravity theories is established by dropping the gravitino 
$(\j^a=0)$ and restricting the auxiliary field $B$ to its real (scalar) component, 
$B=3X$ with $\Bar{X}=X$. Then, as is shown in ref.~\cite{us}, the bosonic 
Lagrangian takes the form
\be \lb{bos}
L = 2F' \left[ \frac{1}{3}R +4X^2 \right] + 6XF
\ee
It follows that the auxiliary field $X$ obeys an algebraic equation of motion,
\be \lb{aux}
3F +  11F'X + F''\left[ \frac{1}{3}R + 4X^2\right] =0
\ee
In those equations $F=F(X)$ and the primes denote the derivatives with respect to
$X$. Solving eq.~(\ref{aux}) for $X$ and substituting the solution back into 
eq. (\ref{bos}) results in the bosonic function $f(R)$. The physical sector
of the $F(\car)$ supergravity is larger than that of the usual supergravity  (ie. 
graviton and gravitino) due to the extra scalar (inflaton), its pseudo-scalar 
superpartner (axion) and inflatino.  

It is natural to expand the input function $F(\car)$ in  power series of $\car$.
For instance, when $F(\car)=f_0 -\frac{1}{2}f_1\car$ with some (non-vanishing and 
complex) coefficients $f_0$ and $f_1$, one recovers the standard {\it pure} N=1 
Poincar\'e supergravity with a negative cosmological term \cite{us,kwata}. A generic
$\car^2$ supergravity and the corresponding $f(R)$ functions were studied in
ref.~\cite{kwata}. The most relevant term for the slow-roll chaotic inflation in
 $F(\car)$ supergravity is {\it cubic} in $\car$. In refs.~\cite{kstar,2kw}
we examined the model with 
\be \lb{cub}
F(\car)= f_0 -\frac{1}{2}f_1 \car + \frac{1}{2}f_2 \car^2 -\frac{1}{6}f_3\car^3
\ee
whose real coupling constants $f_{0,1,2,3}$ are of (mass) dimension $3$, $2$, $1$
and $0$, respectively. The stability conditions (ie. the absence of ghost and 
tachyonic degrees of freedom) require $f_1>0$ and $f_3>0$, whereas the stability of the 
bosonic embedding in $F(\car)$ supergravity requires $F'(X)<0$ \cite{kwata,kstar}. 
For the choice (\ref{cub}) the last condition implies
\be \lb{stab} 
f_2^2 < f_1f_3 
\ee
In ref.~\cite{kstar} we used the stronger conditions
\be \lb{cco}
 f_3 \gg 1~,\qquad f_2^2 \gg f_1 \qquad {\rm and} \qquad  f_2^2\ll f_1f_3 
\ee 
The first condition above is needed to have inflation at the curvatures much less 
than 
$M^2_{\rm Pl}$ (and to meet observations), while the second condition is needed to 
have the scalaron (inflaton) mass be much less than $M_{\rm Pl}$, in order to avoid 
large (gravitational) quantum loop corrections  after the end of inflation up to the
present time. The last condition in eq.~(\ref{cco}) was used in ref.~\cite{kstar} for
simplicity: then the second term on the right-hand-side of eq.~(\ref{cub}) does not 
affect inflation.

Equation (\ref{bos}) with the Ansatz (\ref{cub}) in the case of $f_0=0$ (for simplicity)
reads
\be \lb{bos3}
L = -5f_3X^4 + 11f_2 X^3 - (7f_1 +\frac{1}{3}f_3R)X^2 +\frac{2}{3}f_2RX 
-\frac{1}{3}f_1R
\ee 
and gives rise to a cubic equation on $X$,
\be \lb{aux3} 
X^3 -\left( \fracmm{33f_2}{20f_3}\right)X^2 +\left( \fracmm{7f_1}{10f_3} 
+\fracmm{1}{30}R\right)X - \fracmm{f_2}{30f_3}R =0
\ee
The high curvature regime including inflation  is given by 
\be \lb{reg1}
\d R<0 \quad {\rm and} \quad 
\fracmm{\abs{\d R}}{R_0}\gg \left(\fracmm{f^2_2}{f_1f_3}\right)^{1/3}
\ee
where we have introduced the notation $R_0=21f_1/f_3>0$ and $\d R = R+R_0$. With 
our sign  conventions we have $R<0$ during the de Sitter and matter dominated stages. 
In the regime (\ref{reg1})  the $f_2$-dependent terms in eqs.~(\ref{bos3}) and 
(\ref{aux3}) can be neglected, and we get
\be \lb{aux31}
X^2 = -\frac{1}{30} \d R
\ee
and
\be \lb{lag1}
L = -\fracmm{f_1}{3}R + \fracmm{f_3}{180}(R+R_0)^2
\ee
The value of the coefficient $R_0$ is not important in the high-curvature regime.
In fact, it may be changed by the constant $f_0\neq 0$ in the Ansatz (\ref{cub}). 
Thus eq.~(\ref{lag1}) reproduces the inflationary model  (\ref{star}) since inflation
  occurs at $\abs{R}\gg R_0$.  Therefore, we can identify 
\be \lb{f3}
f_3=\fracmm{15}{M^2} \approx 6\cdot 10^{10}
\ee
where we have used the WMAP-fixed inflaton mass $M$ value \cite{kstar}.

The only significant difference with respect to the original $(R+R^2)$ inflationary 
model is the scalaron mass that becomes much larger than $M$ in supergravity, soon 
after the end of inflation when $\d R$ becomes positive. However, it only makes the 
scalaron decay faster  and creation of the usual matter (reheating) more effective.

The whole series in powers of ${\car}$ may also be considered, instead of the limited
Ansatz (\ref{cub}). The only necessary condition for embedding inflation is that 
$f_3$ should be anomalously large. When the curvature grows, the $\car^3$-term should
 become important much earlier than the convergence radius of the whole series 
without that term. 

The model (\ref{cub}) with a sufficiently small $f_2$ obeying the conditions 
(\ref{cco})  gives a simple (economic) and viable realization of the chaotic 
$(R+R^2)$-type inflation in supergravity, thus overcoming the known (generic) 
difficulty with realization of chaotic inflation in supergravity known as the 
$\eta$-problem \cite{jap1}, without resorting to flat directions and without  
an extra matter superfield with the tuned superpotential needed for stabilization 
of the inflationary trajectory \cite{jap2,kl,lnw}. 

\section{Chaotic inflation in $F(\car)$ supergravity and nonminimal coupling}

Let's now consider the full action (\ref{nmact}) under the slow-roll condition,
ie. when the contribution of the kinetic term is negligible. Then 
eq.~(\ref{nmact}) takes the truly chiral form
\be \lb{slowra}
S_{\rm ch.}= \int d^4x d^2\q \ce\left[ X(\F)\car +W(\F)\right] +{\rm H.c.} 
\ee
When choosing $X$ as the independent chiral superfield, $S_{\rm ch.}$ can
be rewritten to the form
\be \lb{slowra1}
S_{\rm ch.}= \int d^4x d^2\q \ce\left[ X\car -\cz (X)\right] 
+{\rm H.c.} 
\ee
where we have introduced the notation
\be \lb{nota}
\cz(X) = - W(\F(X))
\ee

In its turn, the action (\ref{slowra1}) is equivalent to the chiral $F(\car)$ 
supergravity action (\ref{fact}), whose function $F$ is related to the function 
$\cz$ via Legendre transformation,
\be \lb{ltr}
 \cz= X\car -F~,\quad F'(\car)=X\quad {\rm and}\quad \cz'(X)=\car
\ee
It implies the equivalence between the reduced action (\ref{slowra}) and the 
corresponding $F(\car)$ supergravity whose $F$-function obeys eq.~(\ref{ltr}).

Next, let us consider the special case of eq.~(\ref{slowra}) when the superpotential 
is given by 
\be \lb{wz}
W(\F) = \fracmm{1}{2}m\F^2 +\fracmm{1}{6}\tilde{\l} \F^3
\ee
with the real coupling constants $m>0$ and $\tilde{\l}>0$. The model (\ref{wz}) 
is known as the {\it Wess-Zumino} (WZ) model in 4D, N=1 rigid supersymmetry. 
It has the most general renormalizable scalar superpotential in the 
absence of supergravity. In terms of the field components, it gives rise to the 
Higgs-like scalar potential.

For more simplicity, let's take a cubic superpotential,
\be \lb{wz3}
W_3(\F) = \fracmm{1}{6}\tilde{\l} \F^3
\ee
or just assume that this term dominates in the superpotential (\ref{wz}), and
choose the $X(\F)$-function in eq.~(\ref{slowra}) in the form
\be  \lb{xcho}
 X(\F) = -\x \F^2
\ee
with a large positive coefficient $\x$, $\x>0$ and $\x\gg 1$, in accordance 
with eq.~(\ref{bnma}).

Let's also simplify the $F$-function of eq.~(\ref{cub}) by keeping only the
most relevant cubic term, 
\be \lb{ks3}
F_3(\car) = - \fracmm{1}{6}f_3\car^3
\ee

It is straightforward to calculate the $\cz$-function for the $F$-function
(\ref{ks3}) by using eq.~(\ref{ltr}). We find
\be \lb{eks}
-X =\fracmm{1}{2}f_3\car^2 \quad {\rm and}\quad 
\cz'(X)=\sqrt{\fracmm{-2X}{f_3}}
\ee
Integrating the last equation with respect to $X$  yields
\be \lb{cz}
\cz(X) = -\fracmm{2}{3}\sqrt{\fracmm{2}{f_3}}(-X)^{3/2}=
-\fracmm{2\sqrt{2}}{3}\fracmm{\x^{3/2}}{f^{1/2}_3}\F^3
\ee
where we have used eq.~(\ref{xcho}). In accordance to eq.~(\ref{nota}),
the $F(\car)$-supergravity $\cz$-potential (\ref{cz}) implies the superpotential 
\be \lb{sp}
W_{\rm KS}(\F)= \fracmm{2\sqrt{2}}{3}\fracmm{\x^{3/2}}{f^{1/2}_3}\F^3 
\ee
It coincides with the superpotential (\ref{wz3}) of the WZ-model, 
provided that we identify the couplings as
\be \lb{cid}
 f_3 = \fracmm{32\x^3}{\tilde{\l}^2}
\ee

We thus conclude that the original nonminimally coupled matter-supergravity
theory (\ref{nmact}) in the slow-roll approximation with the superpotential
(\ref{wz3}) is classically equivalent to the $F(\car)$-supergravity theory
with the $F$-function given by eq.~(\ref{ks3}) when the couplings
are related by eq.~(\ref{cid}).  

The inflaton mass $M$ in the supersymmetric case, according to eqs.~(\ref{f3}) 
and (\ref{cid}) is thus given by
\be  \lb{fin}
M^2 = \fracmm{15\tilde{\l}^2}{32\x^3}
\ee
Therefore, according to eqs.~(\ref{ind1}), (\ref{f3}), (\ref{cid}) and (\ref{fin}), 
the value of $\x$ in the supersymmetric case is $\x_{\rm susy}^3=(45/32)\x_{\rm bos}^2$, 
or $\x_{\rm susy}\approx 10^3$. We have assumed here that 
$\tilde{\l}\approx {\cal O}(1)$.

\section{Conclusion}

In this paper we proved the equivalence of two different (viable and well known) 
physical theories of inflation (Higgs inflation vs. $(R+R^2)$ inflation), and extended 
it to supergravity. These results are important because resolution between various 
inflationary theories is one of the main objectives in modern cosmology. It follows 
{}from our results that the CMB alone cannot distinguish between the Higgs and the 
$(R+R^2)$ inflation. However, reheating in both theories is different. In both models 
gravity is modified, so the effective gravitational constant becomes time- and 
(generically) space-dependent, which is of particular importance during inflation. 
However, the physical nature of inflaton in the $f(R)$ gravity and the 
scalar-tensor gravity is very different. In the $f(R)$ gravity the inflaton field 
is the spin-0 part of metric, whereas in the scalar-tensor gravity inflaton is a 
matter particle. The inflaton interactions with {\it other} matter fields are, 
therefore, different in both theories. It gives rise to different inflaton decay 
rates and different reheating, ie. implies different physics after inflation (see 
ref.~\cite{bg1} too). The same remarks apply to the supergravity case. 
Nevertheless, we expect the equivalence between the non-minimally coupled 
supergravity and the $F(\car)$ supergravity to hold during {\it initial} reheating 
with harmonic oscillations. In the bosonic case the equivalence holds until the 
inflaton field value is higher than $M_{\rm Pl}/\x_{\rm bos}\approx 
10^{-5} M_{\rm Pl}$. In the supersymmetric case we find the similar bound at 
$M_{\rm Pl}/\x_{\rm susy}^{3/2}\approx 10^{-5} M_{\rm Pl}$.

It should also be emphasized that our supergravity extension of Higgs inflation is 
truly minimal: it does not have extra symmetries, extra fields and/or extra 
interactions beyond those required by $N=1$ local (Poincar\'e) supersymmetry 
(compare e.g., with the $N=1$ superconformal approach in ref.~\cite{sconfh}).

The established equivalence also begs for a fundamental reason. 
In the high-curvature (inflationary) regime the $R^2$-term dominates over the $R$-term 
in the Starobinsky $f(R)$-gravity function (\ref{star}), while the coupling constant 
in front of the $R^2$-action (\ref{fgr}) is dimensionless. The Higgs slow-roll 
inflation is based 
on the Lagrangian (\ref{bl}), where the $\x\f_J^2$ dominates over $1$  (in fact, over 
$M^2_{\rm Pl}$) in front of the gravitational $R$-term, and the relevant scalar 
potential is given by $V_4=\frac{1}{4}\l\f_J^4$ since the parameter  $v$ is irrelevant 
for inflation, while the coupling constants $\x$ and $\l$ are also  dimensionless. 
Therefore, both actions are {\it scale invariant} in the high field (or high energy)
 limit. Inflation spontaneously breaks that symmetry.

The supersymmetric case is similar: the nonminimal action (\ref{slowra}) with the 
$X$-function (\ref{xcho}) and the superpotential (\ref{wz3}) also has only dimensionless 
coupling constants $\x$ and $\tilde{\l}$, while the same it true for the $F(\car)$-supergravity 
action with the $F$-function (\ref{ks3}), whose coupling constant $f_3$ is 
dimensionless 
too. Therefore, those actions are also asymptotically scale invariant, 
while inflation spontaneously breaks that invariance.

A spontaneous breaking of the scale invariance necessarily leads to Goldstone particle 
(dilaton) associated with the spontaneously broken scale transformations (dilatations). So,
 perhaps, the scalaron (inflaton) of Sec.~3 should be {\it identified} with the Goldstone 
dilaton related to the spontaneously broken scale invariance (dilatations)!

The basic field theory model, describing both inflation {\it and} subsequent 
reheating, reads (see eg., eq.~(6) in the second paper of ref.~\cite{kls})
\be \lb{basm}
\eqalign{
 L/\sqrt{-g}  = &~  \ha \pa_{\m}\f\pa^{\m}\f - V(\f) +\ha \pa_{\m}\c\pa^{\m}\c
-\ha m^2_{\c}\c^2 +\ha \tilde{\x} R\c^2 +\bar{\j}(i\g^{\m}\pa_{\m}-m_{\j})\j \cr
 &~ -\ha g^2\f^2\c^2 - h(\bar{\j}\j)\f \cr}
\ee
with the inflaton scalar field $\f$ interacting with another scalar field $\c$ and
a spinor field $\j$. The nonminimal supergravity (\ref{nmact}) with the  WZ 
superpotential (\ref{wz}) can be considered as the N=1 locally supersymmetric 
extension of the basic model (\ref{basm}), after rescaling $\f_c$ to
$(1/\sqrt{2})\f_c$ and identifying $\tilde{\x}=-\fracm{1}{3}\x$ because of 
eq.~(\ref{sbnm}). Therefore, {\it pre-heating} (ie. the nonperturbative enhancement 
of particle production due to broad parametric resonance \cite{kls}) is a generic feature of
supergravity.

The axion $\c$ and fermion $\j$ are both required by supersymmetry, being in the same
chiral supermultiplet with the inflaton $\f$. The scalar interactions are 
\be \lb{sint}
V_{\rm int}(\f,\c) =m\hat{\l}\f(\f^2 +\c^2) +\fracmm{\hat{\l}^2}{4}(\f^2+\c^2)^2 
\ee  
whereas the Yukawa couplings are given by
\be \lb{yu}
 L_{\rm Yu}= \ha \hat{\l} \f (\bar{\j}\j) +\ha \hat{\l}\c  (\bar{\j}i\g_5\j)  
\ee
Supersymmetry implies the unification of couplings since $h=-\ha \hat{\l}$ and
$g^2=\hat{\l}^2$ in terms of the single coupling constant $\hat{\l}$. If supersymmetry 
is unbroken, the masses of $\f$, $\c$ and $\j$ are all the same. However, inflation
already breaks supersymmetry, so the spontaneously broken supersymmetry is 
appropriate here.

Finally, it can be argued that the classical equivalence is {\it broken} in quantum 
theory because the classical equivalence is achieved via a non-trivial field 
redefinition. When doing that field redefinition in the path integrals defining those 
quantum theories (under their unitarity bounds), it gives rise to a non-trivial Jacobian 
that already implies the {\it quantum nonequivalence}, even before taking into account
renormalization.~\footnote{See ref.~\cite{kmy} for the first steps of quantization 
with a higher time derivative.}

In the supergravity case, there is one more clear reason for the quantum nonequivalence 
between the $F(\car)$ supergravity and the classically equivalent matter (nonminimally) 
coupled supergravity. The K\"ahler potential of the inflaton chiral superfield is described 
by a {\it full} superspace integral and, therefore, receives quantum corrections that can 
easily spoil classical solutions describing an accelerating universe. Actually, it was the 
main reason for introducing flat directions in the K\"ahler potential and realizing slow-roll
inflation in supergravity by the use of a chiral scalar superpotential along the flat 
directions \cite{jap1,jap2}. The $F(\car)$ supergravity 
action is truly {\it chiral}, so that the function $F(\car)$ is already protected against the
quantum perturbative corrections given by full superspace integrals. It explains why we 
consider $F(\car)$ supergravity as the viable and self-consistent alternative to the 
K\"ahler flat directions for realizing slow-roll inflation in supergravity.

\section*{Acknowledgements}

SVK was supported in part by the SFB 676 from the University of Hamburg and DESY (as a 
fellow), by the special Fund of the TMU Faculty of Science, and by the World Premier 
International Research Center Initiative (WPI Initiative, MEXT) in Japan. AAS acknowledges 
the RESCEU hospitality as a visiting professor. AAS was partially supported by the Russian 
Foundation for Basic Research (RFBR) under the grant 11-02-12232-ofi-m-2011. The authors
are grateful to F. Bezrukov, D. Gorbunov, M. Shaposhnikov, S. Theisen and A. Westphal for 
discussions.

\end{document}